# SoaDssPm: A new Service-Oriented Architecture of the decision support system for the Project Management.

Fatima Boumahdi, Rachid Chalal

*Abstract*—This paper presents an architecture for the Project Management, which is defined using the concepts behind Service-Oriented and Decision Support System. The framework described, denominated as SoaDssPm, represents the following: a coherent solution to the problem of control Project Management the existing gap between the real execution of Project Management by describing the business process and relationships required by a SOA solution, and its objectives representation, in which the decisional aspects determine the final shape of the system, providing decision support to the identified business processes and constraints.

## I. Introduction

Time and cost control are the essential management functions for achieving successful delivery of engineering, procurement and construction (EPC) projects. The literature suggests needed improvements in areas related to design, development and implementation of efficient project control system. In addition, there is a need for efficient computational environmental that allows data sharing among members of project teams.

Existing control methods can be classified into two categories. The traditional 'S' curve method uses one cost indicator to compare project's budgeted and actual values. This method tends to mislead the decision-maker: it lacks realistic reporting of cost performance in relation to actual progress made (Moselhi 1993). To overcome this problem, the earned value method (DoD 1967) was introduced, where two indicators (cost and time) were utilized as basis for reporting the progress status of the project. It tracks either cost variance or schedule variance of a project cumulatively at the time of reporting. However, this method neither specifies causes of the calculated variances, nor suggests needed corrective actions (Alshaibani 1999). Clearly, there is a need to develop a new architecture that not only supports the earned value analysis but also identifies the reasons behind calculated variances and suggests corrective actions.

In a few years, service oriented architecture (SOA) became a major topic for the information systems of company [7]. More than one new technology or method, it is the convergence of several existing approaches, and the emergence of a strong data-processing adhesion of the directions and trade with the same objective. SOA is founded on the construction of reusable and flexible services, neutrals compared to the platform of communication and which correspond to the businesses processes of the company.

The decision support systems are present in many fields and aim to help the decision maker in his task by providing him all the relevant elements for decision making. However, the project management adapted in companies do not reflect the reality accurately where various points of view divergent and often conflict must be considered to arrive at a compromise that gave rise to a new dimension: decisional.

This paper presents a proposal that brings together the aforementioned aspects: it aims to define a framework for the Project Management control, which uses the concepts of Service-Oriented for its automatics cooperation. This framework which is denominated as SoaDssPm, comprises an integrated solution that follows the SOA architecture for the Project Management Control automation. Accordingly, and following the separation in real and objectives of Project Management fostered by Decision.

As a result of this idea, our article relates the improvement of the project management by using some knowledge transferred from the branch of decision.

This paper outlines the framework of a newly developed system that aims at circumventing the limitations referred to above and briefly describes the basic components of the system. The proposed system has a number of interesting features: 1) It is a Service Oriented where each service represents an under-process, 2) It provides diagnostic system for causes of poor performance, 3) it forecasts future cost and time at completion, and 4) it provides a real-time data-sharing environment to facilitate the progress reporting to all members of project team. The architecture that allow to reach all these features is SOA, and it's for this purpose that we have used it.

Our article is structured as follows: the following section exposes the problems met in PM, we detail the design of

SoaDssPm in section 3, using MDSE (Model-Based Development with SoaML) method. The detailed architecture of SoaDssPm will be illustrated in section 4. And at last, we finish our article by a conclusion.

## II. PROBLEM STATEMENT

The project management is a highly complex process [6]. For this reason, we cannot give an intuitive description of such project management process. Therefore, we propose a new architecture more accurate and advanced, with the use of DSS and SOA, for its design. In this paper, we propose an architecture that treats the following problems:

### A. Automation

For the realization of a project, much of task and work must be done, and the company must get more worker-time through efficiency drives, training, outsourcing or automation. In our architecture, we want to substitute supplement manual processes to manage information flow within an organization to have lower costs, reduce risk, and increase consistency.

### B. Cooperation

The important aspects of business process cooperation are:
- Causal relations between the main business processes of the company
- Mapping of business processes onto business functions
- Realization of services by business processes
- Use of shared data

### C. Decision

The decision corresponds to the project control, it includes site data collection, performance evaluation and corrective actions. Site data collection involves measuring work progress and calibrating the actual expense of the resources during the construction phase. In performance evaluation, the collected data is compared to the baseline in order to find out any cost overruns or schedule delays. Based on satisfactory results i.e. positive variances, for this evaluation process, the next cycle will be started. Otherwise, further analysis would have to be activated in order to determine the reasons behind the calculated variances and corrective action(s) would be recommended.

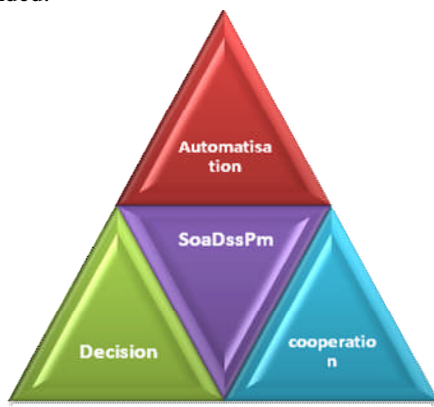

Fig. 1.The problematic of our research

The next section shows the design of ….. to avoid these problems.

## III. MDSE

MDSE (Model-Based Development with SoaML) [4] methodology is chosen to services identification for this architecture.

MDSE methodology aims to integrate with existing business modeling practices within an enterprise, allowing building upon and extending existing modeling practices rather than replacing them [4]. The methodology assumes that modern business modeling practices take advantage of business modeling tools that adopts the OMG MDA specifications Business Motivation Model (BMM) [2] and Business Process Modeling Notation (BPMN) [2].

### A. Business Architecture Model (BAM)

BAM is used to express the business operations and environment which the service-oriented architecture is to support. The BAM includes business goals, business processes with associated organization roles and information elements [4].

Our process breaks up in four sub-process:
- Management of business opportunity :

The business engineer establishes a first contact with the customer and collects all the necessary information for the preparation of the proposal by respecting the customer requirements. The purpose of this stage is to initiate the client relationship and to define the projects like adding them.

- Preparation of the proposal :

The before-sale engineer studies the offer and prepares a proposal by carrying out a design for the solution. The business engineer makes an estimate of the weight of the offer so that Business User can make a decision BID/No BID to validate the proposal or not. This phase aims to prepare the solution to be able to propose it to the customer and to ensure his technical validation.

- Negotiation (business Signature) :

The purpose of this stage is to make a proposal with the customer and to sell the project. The customer consults the proposal, thereafter the before-sale engineer carries out the modifications according to the requirements of the customer, the business engineer evaluates the risks and communicates them to the business manager who must make a Win/Loss decision, thereafter the legal support draws up a contract which must be signed by the customer.

- Implementation of the solution :

It is the beginning of the after-sales phase. Architect makes technical study and creates plan which will be given to the project manager, who in his turn will establish a management plan, he defines the list of the tasks thereafter and assigns those with the foremen. The members of the team carry out the tasks which theirs are assigned. As soon as all the tasks are supplemented, the project manager and his team must test the project. Once all the tests carried out the project will be transmitted to the customer with the statement of receipt and the contract will be closes.

Business Process Modeling Notation (BPMN) [2], is designed to communicate a wide variety of information on business processes to a wide variety of audiences, providing a standard notation that is readily understandable by all business stake-holders. Typically, BPMN is used to define business processes. The next figure shows the modeling of our trade process with BPMN.

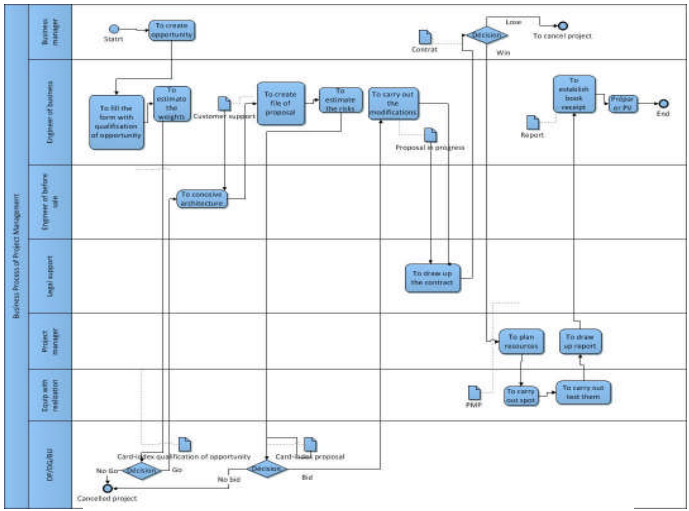

Fig. 2. BPMN of Project Management

- Software Architecture Mode (SAM):

SAM specifies the interfaces and message types, services interfaces, behavior, components and ports [4].

*B. Services Architecture*

Service architecture is a high level description of how participants work together for a purpose by providing and using services expressed as service contracts [3]. The service architecture defines the requirements for the types of participants and service realizations that fulfill specific roles. The next figure shows the services architectures of our process.

The following figure shows the Service Architecture diagram of SoaML.

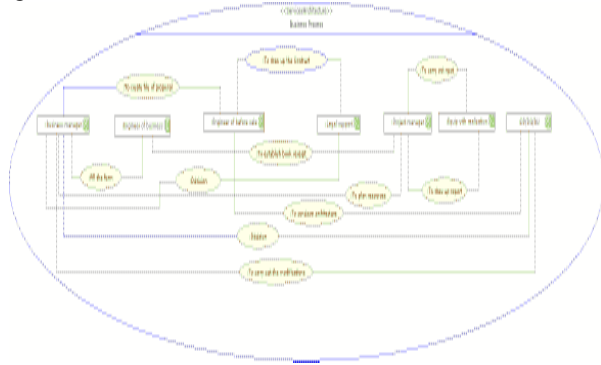

Fig. 3. Service Architecture diagram of Project Management.

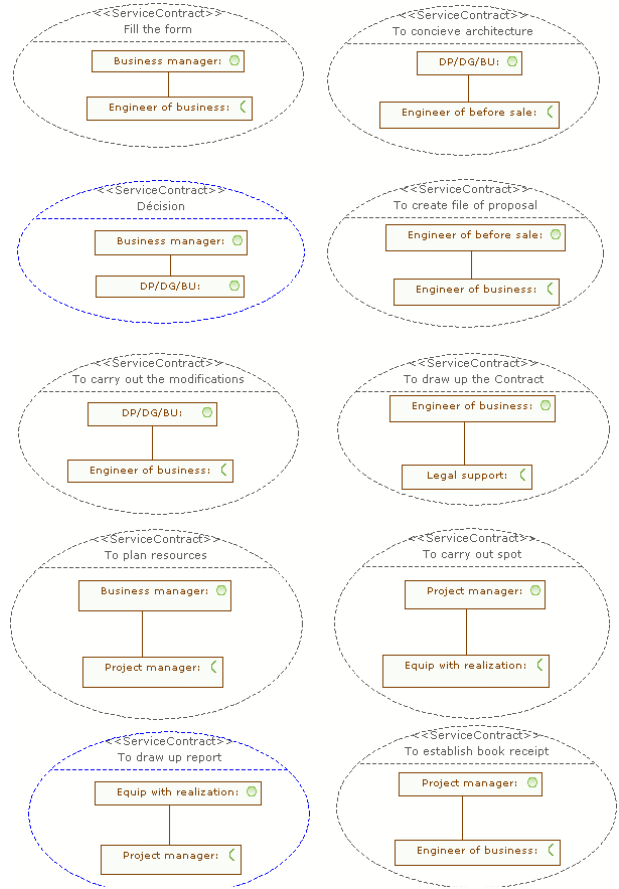

Fig. 4. Service Contract Diagram of Project Management.

IV. ARCHITECTURE

Below is a brief describing of the main components of the system.

*A. Collaborative System allowing :*

- To ensure the individuals the automation of the individual or collective tasks by the installation of *workflows* within an integrated collaborative space (document management, transport, instantaneous communication…);
- To automate the processes by integrating the applications and partners;
- To manage complex trades rules;

*B. Decision Support System*

The principal role of DSS is to control a project effectively, to reach this objective, we use earned value analysis method to design our DSS.

Earned value analysis (EVA) [1] is a method of performance measurement. Many project managers manage their project performance by comparing planned to actual results. With this method, one could easily be on time but overspend according to the plan. A better method is earned value because it integrates cost, schedule and scope and can be used to forecast future performance and project completion dates. It is an "early warning" program/project management tool that enables managers to identify and control problems before they become insurmountable. It allows projects to be managed better on time, on budget. Following is the summary of important Earned Value terms and formula.

TABLE I
EARNED VALUE MANAGEMENT TERMS

| Term | Description | Interpretation |
|---|---|---|
| PV (BCWS) | Planned Value | What is the estimated value of the work planned to be done? |
| EV (BCWP) | Earned Value | What is the estimated value of the work actually accomplished? |
| AC (ACWP) | Actual Cost | What is the actual cost incurred? |
| BAC | Budget at Completion | How much did you BUDGET for the TOTAL JOB? |
| EAC | Estimate at Completion | What do we currently expect the TOTAL project cost? |
| ETC | Estimate to Complete | From this point on, how much MORE do we expect it to cost to finish the job? |
| VAC | Variance at Completion | How much over or under budget do we expect to be? |

TABLE II
EARNED VALUE MANAGEMENT FORMULA AND INTERPRETATION

| Name | Formula | Interpretation |
|---|---|---|
| Cost Variance (CV) | EV – AC | NEGATIVE is over budget, POSITIVE is under budget |
| Schedule Variance (SV) | EV – PV | NEGATIVE is behind schedule, POSITIVE is ahead of schedule |
| Cost Performance Index (CPI) | EV / AC | I am [only] getting ______ vents out of every $1. |
| Schedule Performance Index (SPI) | EV / PV | I am [only] processing at ______ % of the rate originally planned. |
| Estimate At Completion (EAC) **Note**: There are many ways to calculate EAC. | BAC / CPI<br><br>AC + ETC<br><br>AC + BAC – EV<br><br>AC + (BAC – EV) / CPI | As of now how much do we expect the total project to cost $ _____.<br>• Used if no variances from the BAC have occurred<br>• Actual plus a new estimate for remaining work. Used when original estimate was fundamentally flawed.<br>• Actual to date plus remaining budget. Used when current variances are atypical.<br>• Actual to date plus remaining budget modified by performance. When current variances are typical. |
| Estimate To Complete (ETC) | EAC – AC | How much more will the project cost? |
| Variance At Completion (VAC) | BAC – EAC | How much over budget will we be at the end of the project? |

*C. SOA Architecture*

Our ArchiGpDes architecture illustrated by fig.4 is made up of four layers in accordance with the IBM model [7]. The SoaDssPm must be able to make it possible to define specific indices to each company and to build indicators adapted to the needs for the company. The SoaDssPm is thus separated in four layers (fig.4), as follows:

- The Layer 1 "**Data**": it contains two under layers; the first is "Trade" which includes the services trade (CRUD) business process (Project Management) carried out in the company. The second under layer is the "Indices", its goal is to safeguard the indices services of EVA method.

- The Layer 2 "**Technique**": it contains two under layers: under layer "Function" which represents the function services of the process business (Project Management). Under layer "Models", which included model services of EVA method.

- The Layer 3 "**Action**": it contains two under layers: Applicative and Indicators. The first gathers the Applicative services of the process trade (Project Management), and the second ensures the appreciation of the indicators, using the EVA method, and a mode of visualization.

- The Layer 4 "**Presentation**": it contains interfaces, and ensures the communication between the user and the system, that is to say to carry out the process trade, or to make a decision.

Moreover, we conceived our approach so that it respects IDC model of SIMON [5]. We have followed what is perhaps the most widely accepted categorization of the decision-making process first introduced by Herbert Simon. Simon's categorization of the decision-making process consists of three phases (IDC): Intelligence, Design, and Choice. On the level of the presentation layer, the decision maker can make a decision according to the indicators, in more the mode of visualization helps the decision maker in his choice. In the model layer, the platform makes it possible to conceive the solution (indicators). The purpose of the last layer "indices" is to identify and formulate the problem.

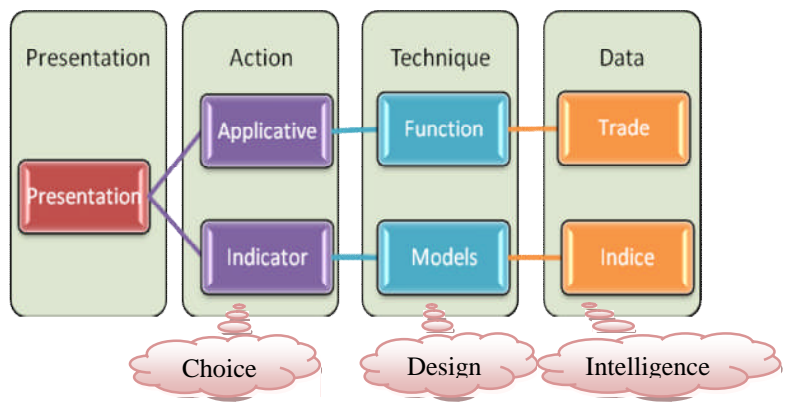

Fig. 6. SoaDssPm architecture

The following figure illustrates the detailed architecture of SoaDssPm.

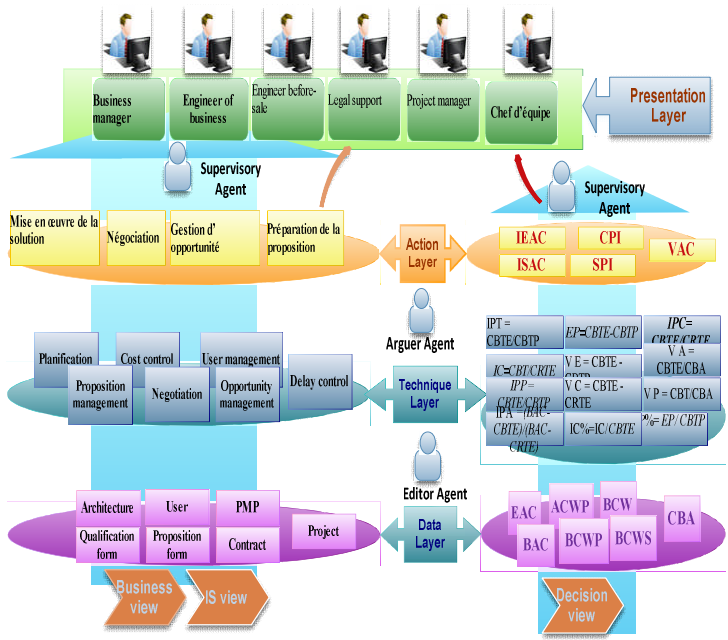

**Fig. 7.** The detailed architecture of SoaDssPm

## V. CONCLUSIONS AND OUTLINES

At the end of this article, we proposed an SOA decisional architecture for project management.

In this work, a service oriented architecture and Decision support system allowing the resolution of problem of PM aid were presented. It is made up of a user interface, Action layers, Technique and of a Data layer, each layer is made up of two under layers, to represent the process trade, other intended for the decision-making process. These various layers allow to the user to carry out his process trade, and solve his problem according to his needs. The suggested architecture assists the user throughout his decision-making process without substituting him.

Thus, we proposed the definition of new service types intended to carry out the decision-making aid. Three service types were proposed:

- Services of indices, intended to represent the general framework, the context in which the decision is carried out,
- Services of model, allowing to identify the rules of decision,
- Services of indicator, intended to indicate the waiting of the decision maker,

In our future work, we envisage the enrichment of our architecture (SoaDssPm) to which we will add new modules and new classes which will allow to model the real systems more easily and to develop other computing systems of decision-making aid.

Like prospect, there remains to us the implementation of a framework suggested architecture, and to test it on different fields. We also think to use another methods and tools of decision-making aid like the multi criterion methods.